\documentclass[11pt,twoside]{article}
\usepackage{./asp2014}

\aspSuppressVolSlug
\resetcounters

\begin{document}

\title{Understanding the spectrum of the very hot DA white dwarf PG0948+534}
\author{S. P. Preval,$^1$ and M. A. Barstow,$^2$
\affil{$^1$University of Strathclyde, Department of Physics, Glasgow G4 0NG UK; \email{simon.preval@strath.ac.uk}}
\affil{$^2$University of Leicester, Department of Physics and Astronomy, University Road, Leicester, LE1 7RH UK; \email{mab@le.ac.uk}}}

\paperauthor{S. P. Preval}{simon.preval@strath.ac.uk}{ORCID_Or_Blank}{University of Strathclyde}{Department of Physics}{Glasgow}{Lanarkshire}{G4 0NG}{UK}
\paperauthor{M. A. Barstow}{mab@le.ac.uk}{ORCID_Or_Blank}{University of Leicester}{Department of Physics and Astronomy}{Leicester}{Leicestershire}{LE1 7RH}{UK}

\begin{abstract}
PG0948+534 is currently one of the hottest DA white dwarf stars, and is also one of the most mysterious. Attempts to model the sharp, deep absorption features of this star have been unsuccessful. In these proceedings we describe our analysis of PG0948+534. We perform a line survey of the UV spectrum of PG0948+534, making detections of 300+ absorption features, and identifying four distinct velocity regimes. We find evidence of circumstellar absorption in the profiles of C {\sc iv} and Si {\sc iv}. Using non-local thermodynamic equilibrium model atmospheres, we are able to correctly model the absorption features of the star, providing abundance measurements for C, N, O, and Si for PG0948+534 for the first time. We also revise the effective temperature and gravity for this star using models including these new abundances.
\end{abstract}

\section{Introduction}
PG0948+534 is one of the hottest known DA white dwarf stars. PG0948+534 is also interesting theoretically due to difficulties in modelling the absorption spectrum of the star. \cite{barstow03b} attempted to measure the effective temperature ($T_{\mathrm{eff}}$), surface gravity (log $g$), and metal abundances for this star. They calculated a non-local thermodynamic equilibrium (NLTE) metal-polluted model grid spanning a wide range of $T_{\mathrm{eff}}$ and log $g$. To model the abundances, the grid was calculated for a fixed set of abundances based on a prior analysis of G191-B2B, but other abundance values were calculated in LTE by stepping away from the original model. The authors found that while they could measure abundances for Fe and Ni, it was not possible to measure abundances for C, N, O, and Si. This was because their synthetic model profiles were pressure broadened far beyond the level observed, and did not descend deeply enough into the continuum. The authors concluded that the atmospheric distributions of C, N, O, and Si might be stratified in some way. Despite these shortcomings, the authors were able to measure $T_{\mathrm{eff}}$ and log $g$ for PG0948+534, finding $T_{\mathrm{eff}}=110,000$K and log $g=7.58$. 

Until the present work, study of this object has waned. Using detailed NLTE models, we have been able to model the absorption profiles of C, N, O, and Si, providing abundances for these species for the first time. In these proceedings, we present our current progress in analysing the atmosphere of PG0948+534. We begin by describing the observational data that was used. Next, we discuss the line survey performed, detailing absorption feature parameterisation, and the species detected. Then, we discuss the model atmospheres that were calculated, and how we measured the metal abundances, $T_{\mathrm{eff}}$, and log $g$ of PG0948+534. We then present our results, and compare them to measurements made by other authors. Lastly, we discuss the fitting issues described by \cite{barstow03b} and provide a remedy for them. We finish with a plan for future work, and concluding remarks.

\section{Observations}

We used coadded UV spectra from the {\textit Far Ultraviolet Spectroscopic Explorer} (FUSE), and the {\textit Space Telescope Imaging Spectrometer} (STIS) aboard the Hubble Space Telescope. The FUSE spectrum consists of four observations taken with the LWRS aperture spanning 910-1185\AA. The STIS spectrum consists of four observations taken with the E140M grating spanning 1160-1700\AA. The FUSE and STIS observations are summarised in Table \ref{table:obs}. For both FUSE and STIS, the datasets were coadded, weighted by exposure time. The optical spectrum of PG0948+534 was taken using the Steward Observatory's 2.3m telescope equipped with a Boller \& Chivens spectrograph. This provided a spectrum covering 3400-5600\AA\, with an FWHM of $\sim{6}$\AA. 

\begin{table}
\centering
\caption{List of observations used in constructing the UV spectra.}
\begin{tabular}{@{}lcccc}
\hline
Obs ID & Telescope & Aperture/Grating & Obs date & Exp time (s) \\
\hline
A03415010 & FUSE & LWRS  & 03/11/2001 & 5363.0 \\
A03415020 & FUSE & LWRS  & 06/04/2002 & 6572.0 \\
A03415030 & FUSE & LWRS  & 25/11/2002 & 4283.0 \\
O59P05010 & HST  & E140M & 20/04/2000 & 2628.0 \\
O59P05020 & HST  & E140M & 20/04/2000 & 3110.0 \\
O59P05030 & HST  & E140M & 20/04/2000 & 3110.0 \\
O59P05040 & HST  & E140M & 20/04/2000 & 3110.0 \\
\hline
\label{table:obs}
\end{tabular}
\end{table}

\section{Method}
\subsection{Absorption feature quantification}
We performed a survey of the absorption features present in the UV spectra of PG0948+534. This was done by fitting a Gaussian profile to the absorption features parameterised by the centroid wavelength, line width, and the line strength. In our survey, we were able to resolve four distinct velocity regimes. The velocity for each regime was calculated using absorption features measured in the STIS spectrum. The absorption features were first identified, and their velocities measured. Then, for each regime, the average of these velocities was calculated weighted by the inverse square of their uncertainties. The FUSE spectrum could not be used for this purpose as the absolute velocity calibration is typically very poor (see \citealt{moos00a}), however, relative velocity measurements can be used.

\subsection{$T_{\mathrm{eff}}$, log $g$, and abundance measurement}
All model atmospheres in this work were calculated using the non-LTE (NLTE) model atmosphere code {\sc tlusty} \cite{hubeny88a,hubeny95a}, and synthesised with {\sc synspec} \citep{hubeny11a}. The H Lyman and Balmer line profiles were calculated using the Stark broadening tables of \cite{tremblay09a}. 

To measure the abundances, we used a similar method to that described in \cite{preval13a}, which we briefly describe here. A model atmosphere was first calculated in NLTE with $T_{\mathrm{eff}}$=110,000K, log $g$=7.58, and metal abundances C/H=$1.00\times{10}^{-6}$, N/H=$1.00\times{10}^{-6}$, O/H=$1.00\times{10}^{-5}$, Si/H=$1.00\times{10}^{-6}$, P/H=$1.00\times{10}^{-7}$, S/H=$1.00\times{10}^{-7}$, Ar/H=$3.00\times{10}^{-5}$, Fe/H=$1.00\times{10}^{-6}$, and Ni/H=$1.00\times{10}^{-6}$ as number fractions. The He abundance was fixed to $1.00\times{10}^{-7}$. For each metal, we used {\sc synspec} to step away from this model in LTE to calculate a grid of metal abundances. Finally, we used XSPEC \citep{arnaud96a} to interpolate these model abundance grids to observed data to find the best fitting abundance values.

With the measured abundances, we then calculated a model grid in NLTE spanning $100,000<T_{\mathrm{eff}}<200,000$ in steps of 5000K, and $7.00<\mathrm{log}\,g<9.00$ in steps of 0.50 dex. We synthesized two spectra for each model grid point; one covering the H lyman series, and one covering the H Balmer series. In addition, each spectrum was synthesized using the Stark broadening tables of either \cite{lemke97a} or \cite{tremblay09a}. We then used XSPEC to interpolate these models to observe H Lyman and Balmer lines to extract $T_{\mathrm{eff}}$ and log $g$.

\section{Results}

\subsection{Line survey}
The signal to noise of both the FUSE and STIS spectra is poor, making it difficult to resolve small absorption features. However, we were able to resolve 300+ lines in our survey. From the photosphere, we made identifications of C, N, O, Si, P, S, Ar, Fe, and Ni. We resolved four velocity regimes, one pertaining to the photosphere, and three pertaining to the ISM. We measured the velocity of the photospheric component to be $-10.9\pm{0.18}$km s$^{-1}$, and the three interstellar components to be $-17.4\pm{0.39}$, $1.00\pm{0.44}$, and $20.6\pm{0.38}$km s$^{-1}$ respectively. 

We found that it was possible to resolve some absorption features into multiple components. For example, the C {\sc iv} 1550.772\AA\, absorption feature could be split into photospheric and circumstellar components with velocities of $-11.62\pm{2.45}$ and $-17.20\pm{0.87}$ km s$^{-1}$ respectively. Likewise, the Si {\sc iv} 1393.755 absorption feature could also be split into photospheric and circumstellar components with velocities of $-11.48\pm{2.73}$ and $-17.51\pm{1.07}$ km s$^{-1}$ respectively. The circumstellar velocity component is strikingly similar to the velocity of one of the interstellar components mentioned above. The similarity of these velocities suggests that the circumstellar component could arise due to Stromgren sphere ionization as discussed by \cite{dickinson12b}.

\subsection{Parameter measurement}
In Table \ref{table:abns} we list the transitions used to measure the metal abundances, and the abundances measured, including C, N, O, and Si for the first time. Our measurement of the Ar abundance corroborates the value measured by \cite{werner2007a}. We also find large differences between our Fe/Ni abundances compared to \cite{barstow03b}, who measured Fe/H= $1.90\times{10}^{-6}$ and Ni/H = $1.20\times{10}^{-7}$. This may be due to the composition of our initial model atmosphere, which includes P, S, Ar, and much higher abundances compared to the model used by \cite{barstow03b}.

\begin{table}
\centering
\caption{Abundances measured for PG0948+534, and the transitions used to measure them.}
\begin{tabular}{@{}llc}
\hline
Metal & Transition (\AA) & Measured X/H \\ 
\hline
C {\sc iv}   &  1548.187, 1550.772                                   &  $2.83_{-2.03}^{+10.72}\times{10}^{-6}$ \\
N {\sc v}    &  1238.821, 1242.804                                   &  $9.08_{-0.64}^{+0.64}\times{10}^{-7}$ \\
O {\sc v}    &  1371.296                                                    &  $1.51_{-0.28}^{+0.28}\times{10}^{-5}$ \\
Si {\sc iv}  &  1066.614, 1393.755, 1402.770                   &  $1.54_{-0.17}^{+0.17}\times{10}^{-5}$ \\
P {\sc v}    &  1117.977, 1128.008                                     &  $9.24_{-1.51}^{+3.04}\times{10}^{-8}$ \\
S {\sc vi}   &  933.378, 944.523                                        &  $1.37_{-0.48}^{+0.49}\times{10}^{-6}$ \\
Ar {\sc vii} &  1063.630                                                     &  $1.08_{-0.12}^{+0.33}\times{10}^{-6}$ \\
Fe {\sc vi}  &  1337.793, 1370.734, 1371.073, 1374.629 &  $1.24_{-0.25}^{+0.25}\times{10}^{-5}$ \\
Ni {\sc v}   &  1306.624, 1318.515, 1336.136                   &  $3.81_{-1.06}^{+1.18}\times{10}^{-6}$ \\
\hline
\label{table:abns}
\end{tabular}
\end{table}

We measured $T_{\mathrm{eff}}$ and log $g$ for the Lyman and Balmer lines using models that were synthesised using either the Lemke or Tremblay stark broadening tables. When using the Lyman lines, we measure $T_{\mathrm{eff}}=138,000_{-6000}^{+7000}$K and log $g=8.43_{-0.09}^{+0.10}$ for the Lemke tables, and $T_{\mathrm{eff}}=140,000_{-6000}^{+8000}$K and log $g=8.60_{-0.10}^{+0.09}$ for the Tremblay tables. When using the Balmer lines, we measure $T_{\mathrm{eff}}=138,000_{-9000}^{+11000}$ and log $g=7.42_{-0.13}^{+0.16}$ for the Lemke tables, and $T_{\mathrm{eff}}=141,000_{-9000}^{+12000}$K and log $g=7.46_{-0.14}^{+0.18}$ for the Tremblay tables. While the temperatures are consistent over broadening tables and line series, a large discrepancy is seen with the measured log $g$. It is not immediately obvious why this difference arises. 

\section{Resolving the modelling issue}
As mentioned in the introduction, \cite{barstow03b} attempted to model the photospheric absorption features of PG0948+534, but found that the model profiles for C, N, O, and Si were pressure broadened far beyond the level observed in the STIS spectrum, and did not descend deeply enough into the continuum. \cite{barstow03b} used a model grid that spanned a wide range of $T_{\mathrm{eff}}$ and log $g$, and was calculated setting C/H=$4.00\times{10}^{-7}$, N/H=$1.60\times{10}^{-7}$, O/H=$9.60\times{10}^{-7}$, Si/H=$3.00\times{10}^{-7}$, Fe/H=$1.00\times{10}^{-5}$, and Ni/H=$5.00\times{10}^{-7}$. The authors then used {\sc synspec} to step away in LTE to create a grid of metal abundances to compare with observation.

We attempted to reproduce this work by calculating a model atmosphere with the above abundances, $T_{\mathrm{eff}}$=110,000K, and log $g$=7.58. We then used {\sc synspec} to vary the metal abundances in LTE.
As mentioned above, C {\sc iv} and Si {\sc iv} can be modelled by including an additional absorber. We found that we could successfully model the N {\sc v} and O {\sc v} profiles using this model atmosphere.

To aid convergence, {\sc tlusty} is able to treat a set number of levels in detailed balance especially in the case where a transition is very strong. As a test, we recalculated the model atmosphere with the same parameters as before, but treated the ground and first excited level of N {\sc v} in detailed balance. Again, we used {\sc synspec} to vary the abundances. We found that the synthetic N {\sc v} profile no longer matched the observed profile. In Figure \ref{fig:nvfits} we have plotted our attempts to reproduce the observed N {\sc v} profiles using models that were calculated when treating N {\sc v} in NLTE, and when treating the first two levels of N {\sc v} in detailed balance. It can be seen in the detailed balance case that by increasing the N abundance, the synthetic profile becomes broader and a re-emission core begins to form, flattening the profile. Aside from aiding convergence in model atmosphere calculations, detailed balance can be assumed when the metal abundance is small. Therefore, it appears that the issues in modelling the observed metal profiles by \cite{barstow03b} was caused by using models outside their range of validity, whilst also not including the necessary NLTE physics in their model calculations. 

\articlefiguretwo{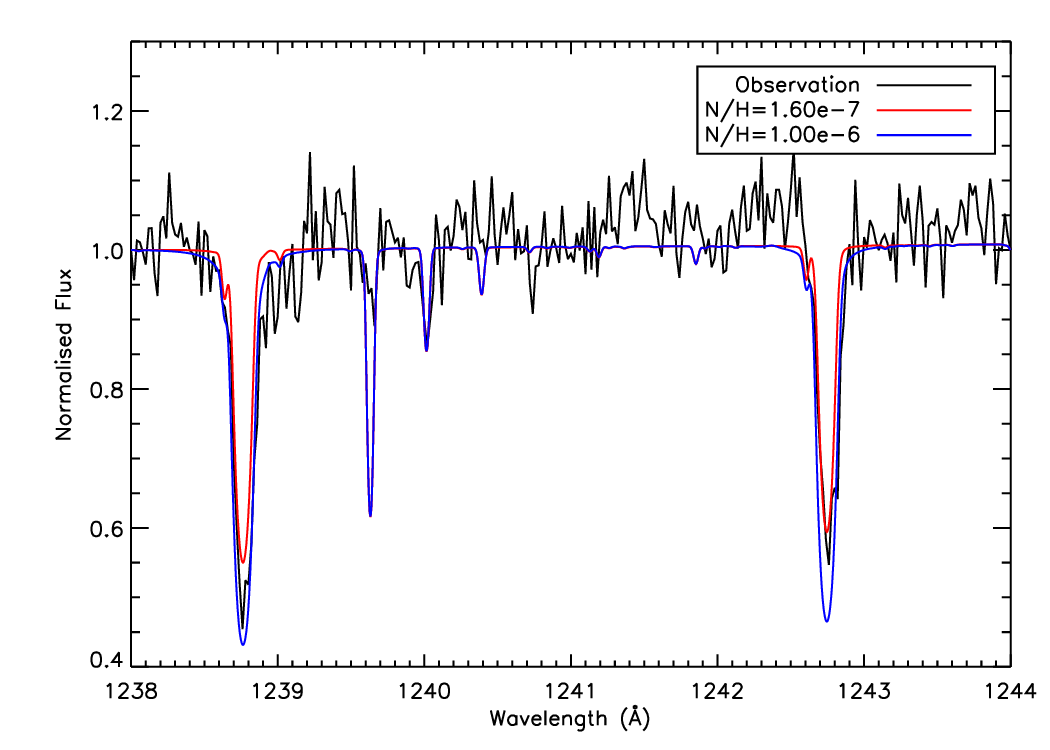}{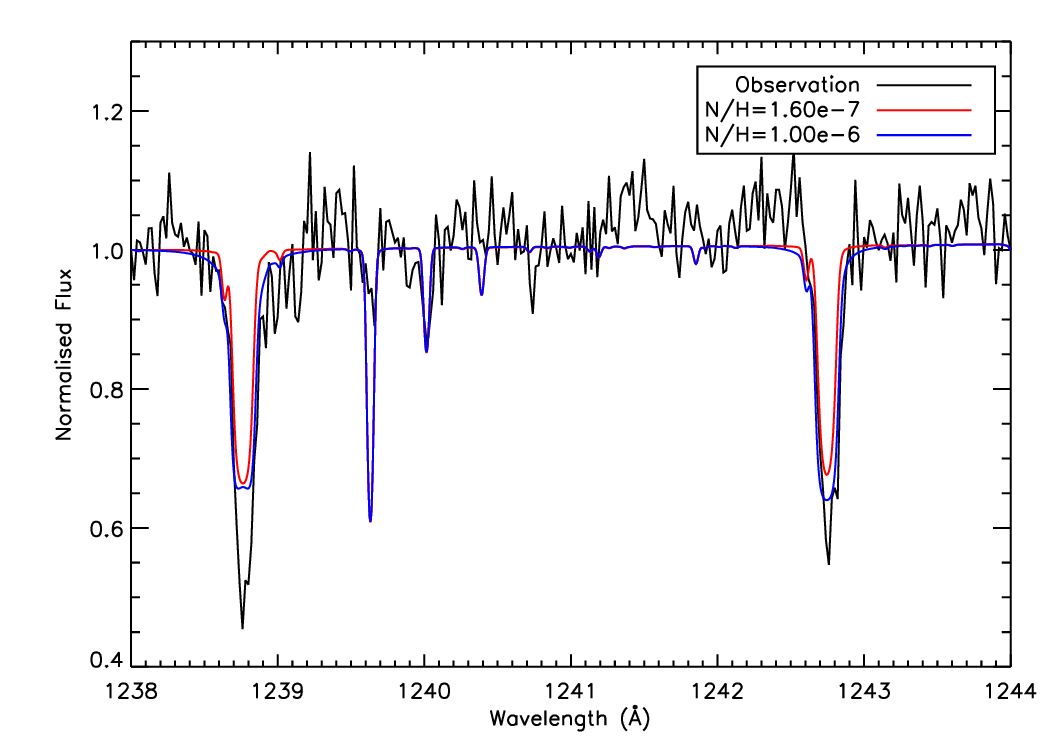}{fig:nvfits}{Plot of observed N {\sc v} 1238 and 1243\AA\, profiles, and synthetic profiles with abundances of N/H=$1.60\times{10}^{-7}$ (red line) and N/H=$1.00\times{10}^{-6}$ (blue line). On the left plot, the synthetic models treat all N {\sc v} levels in NLTE, whereas the right plot shows synthetic models where the first two levels of N {\sc v} were treated in detailed balance.}

\section{Future Work}
The difference between the previous and current models is very large in terms of the measured metal abundances, $T_{\mathrm{eff}}$, and log $g$. Because we have measured the abundances in LTE, and hence  have not accounted for changes to $T_{\mathrm{eff}}$ and log $g$, it is likely that the measurements we have made are not the true values. This may also be the reason that there is such a large discrepancy between log $g$ measured using either the H Lyman or Balmer lines. In order to find the true values, the abundances, $T_{\mathrm{eff}}$, and log $g$ need to be measured in an iterative fashion until convergence is achieved.

\section{Conclusion}
We have summarised our progress thus far on our analysis of the very hot DA white dwarf PG0948+534. We have found that previous attempts to model the photospheric absorption features of this object failed due to the model grid used. While variations in $T_{\mathrm{eff}}$ and log $g$ for this grid were calculated assuming NLTE, abundance variations were calculated assuming LTE due to the restrictive computer speeds available at the time. We performed a line survey and were able to resolved over 300 absorption features. From this survey, we were able to resolve four velocity components, one pertaining to the photosphere, and three pertaining to the ISM. We also found evidence of circumstellar absorption in C and Si. We measured the abundances of several metals in the spectrum of PG0948+534. We re-measured $T_{\mathrm{eff}}$ and log $g$ using the H Lyman and Balmer lines. We found that the Lyman and Balmer $T_{\mathrm{eff}}$ agree reasonably well, and are 30,000K larger than the previous measurement given in \citet{barstow03b}. Furthermore, we noted a remarkable difference between log $g$ measured using the Lyman and Balmer lines, differing by $\sim{1}$ dex. We plan to re-measure the metal abundances, $T_{\mathrm{eff}}$, and log $g$ in an iterative fashion improve the quality of our models, and also to see if this improves the discrepancy between log $g$ values measured using the H Lyman or Balmer lines.

\acknowledgements We are grateful to Pierre Bergeron for kindly providing the optical spectrum used in this work.

\end{document}